# Effects of disorder on superlensing in two dimensional photonic crystal slabs

**X. Wang and K. Kempa**

*Department of Physics, Boston College, Chestnut Hill, Massachusetts 02467*

**Abstract**

We demonstrate that a hexagonal two dimensional photonic crystal can act as an isotropic medium with an effective refractive index $n_{eff} = -1$, therefore capable of unrestricted superlensing. We study the superlensing by calculating the photonic band structure, and the propagation maps of the electromagnetic waves through slabs of the photonic crystal. We investigate the surface, and the crystal disorder effects on the superlensing, by analyzing the light intensity profile at the image of a line source.

wangxb@bc.edu    Sept. 19, 2004

Left-handed materials (LHM), initially proposed by Veselago 36 years ago [1], have recently attracted strong research interests, since their unusual properties can now be realized in the nanoscopic materials [2-13, 29]. For electromagnetic waves incident from air on a slab of LHM, the angle of refraction is negative, which implies via Snell's Law, a negative refractive index. The most exciting consequence of this, is the possibility of the superlensing [13]. A slab made of a material with an isotropic refractive index $n = -1$ restores not only phases of the transmitted propagating waves, but also amplitudes of the evanescent waves that are responsible for the subwavelength details of the source geometry [13]. Such a material (metamaterial) can be used to make a superlens capable of the subwavelength imaging. Limitations of this perfect imaging, resulting from the inherent losses in the media, has also been investigated [14-15]. Various approaches were proposed to fabricate systems, which simulate LHM. In one approach, a model metamaterial was made of the split-ring resonators, and a network of wires [3-5]. Recently, several authors have examined other types of possible structures, which could lead to LHM at infrared and visible frequencies [16-18].

In another approach, it has been shown that a dielectric, two-dimensional (2D) photonic crystal (PC) can act as a LHM, with an effective negative refractive index [19]. It was also demonstrated [20-21], that a negative refraction can be achieved in a PC in which an isotropic refractive index cannot be defined. In this case, an equi-frequency contour in the band structure is chosen so that the group velocities of photon modes excited by an incident wave on this contour, point always in a negative-refraction direction. This all-angle negative refraction can lead to superlensing, but the imaging is severely restricted [22]. Recently, we have demonstrated an unrestricted superlensing in a hexagonal 2DPC slab, and show that the refraction of light follows simple rules of the geometric optics with the Snell's law refraction at each interface, and an effective isotropic refractive index $n_{eff} = -1$ for light propagating inside the crystal [23]. Thus, this PC is one of the most promising model systems for a LHM, and a superlens, and we study here effects of the surface symmetry, and the crystal disorder on the superlensing in this crystal.

The PC systems studied here consist of a hexagonal lattice (in the x-y plane) of infinitely long, cylindrical (in the z-direction) air-holes with a lattice constant $a$ and a hole radius $r = 0.4a$ in a dielectric matrix with $\varepsilon = 12.96$. Each slab is assumed to have a finite width in the x-direction, but is unrestricted in the y-direction. Only the transverse magnetic modes (the electric field parallel to cylindrical holes) are considered. In our simulations, the standard plane wave expansion method [24] is employed to obtain the photonic band structures. We have found that employing 500 plane waves in our calculation assures convergence. To obtain maps of the wave propagation we use the finite-difference time-domain (FDTD) simulations [25-26], with the perfectly matched layer boundary condition [27] in the x-direction, and the periodic boundary condition in the y-direction. The FDTD method discretizes space and time by replacing the partial derivatives in Maxwell's equations with the centered finite differences. The electric and magnetic fields are defined on two spatially interleaved grids. The dielectric permittivity and the magnetic permeability of the structure, and its surrounding medium, are defined at the grid points. Each unit cell contains 1600 ($40 \times 40$) grid points, which guarantees good convergence. The total number of the time steps is 100,000, with each time step of $\Delta t = 0.95/(c\sqrt{\Delta x^{-2} + \Delta y^{-2}})$, where $c$ is the speed of light, and $\Delta x, \Delta y$ are space intervals in the $x$ and $y$ directions, respectively.

Fig. 1(a) shows the normalized frequency $\omega = a/\lambda$ versus the wave vector $k$ in the first Brillouin Zone of the hexagonal 2DPC. The inset shows the corresponding equi-frequency contour, obtained by crossing the second band with the constant frequency ($\omega_0$) plane. The essentially circular shape of this equi-frequency contour indicates that an isotropic effective refractive index of this PC, at this particular frequency, can be defined [19, 28-29]. The equi-frequency contour can have different shapes for various frequencies. From this shape we can deduce the propagation angle inside the PC using the relation $v_g = \nabla_k \omega$. Finally, we can define an effective refractive index using the Snell's law. Fig. 1(b) shows the effective refractive index $n_{eff}$ versus the normalized frequency. Note, that at $\omega_0 = 0.31$ $n_{eff} = -1$, and the equi-frequency contour is circular (Inset in Fig. 1a). These are the conditions for an unrestricted superlensing.

Recently, we have studied in detail superlensing in this PC [23]. A LHM with $n = -1$ perfectly matches to free space [13], and therefore its should not reflect any incoming waves. However, a PC is a model LHM, which behavior is determined by its coarsity, i.e. dimensions of its building blocks (rods, air holes, lattice constant). Only if those dimensions are much smaller than the wavelength of the radiation $\lambda$, a perfect LHM behavior is expected. Since this condition is very difficult to fulfill, we investigate here the consequences of the limited coarsity of the PC. These will include, amongst others, a limited resolution, back reflections, and sensitivity to the surface symmetry.

To test this, we consider an imaging system composed of a PC slab with nine rows of air holes. Fig.2 shows the calculated electric field patterns, generated (at a steady state) by a line source (along z-direction) located to the left of the slab at the centers of the dark blue spots, for three different slabs. The vertical scale bar, to the right, quantifies the relative intensity. The slab thickness is $7.8a$ in Fig. 2(a), and $7.26a$ in Figs.2(b) and (c), corresponding to $4.8\lambda$ and $4.5\lambda$ respectively. While the slabs shown in Figs. 2(a) and (b) possess mirror symmetry with respect to the vertical line through the slab center, the one shown in Fig. 2(c) does not. The corresponding field intensity distributions, plotted along the image center parallel to the slab surface, are shown in Fig.3. The spatial resolution of the imaging, defined as the ratio of the full width at half maximum to $\lambda$, is $R = 0.27$ in all cases. However, the image strength is maximized if the slab thickness is equal to an odd-integer multiple of $\lambda/4$, and the slab possesses mirror symmetry, the case shown in Fig.2(b). At this condition, the reflections of the partial waves generated by the source, are minimized. This conclusion can also be reached by qualitative inspection of the images in Fig.2. Back reflections, represented by additional nodes of intensity and deviations of the wave profiles from circles on the left (source) side of the pictures, are clearly visible in Figs.2(c), and especially 2(a), and less so in Fig.2(b).

The field intensity patterns shown in Fig.2 are dominated outside the slab by single cylindrical waves, with some standing wave contributions due to back reflections. Superlensing implies amplification, inside of the slab, of the evanescent filed components, which normally limit the resolution [13]. While these weak evanescent components do exist in our system, they are difficult to observe in Fig.2, as they are overwhelmed by the complex pattern of standing and propagating waves inside of the slab.

We now turn into the effects of disorder on the superlensing. These are to be expected, since the self-assembly process, which is the favorite nanostructure microfabrication technique, is unlikely to yield perfectly periodic structures. For example, there will be some random deviations of the cylinder (or air-hole) radii, and the inter-cylinder (air-hole) separation from the designed values, and this could lead to image distortion. We consider here the position and the size kind of disorder. For the position disorder, the cylinder positions are randomized, while their radii remain fixed. For the size disorder, cylinder centers remain at lattice sites, but their radii are randomized. To create a random position configuration, we move each cylinder from its lattice position by a distance $\Delta d_x^i = \gamma_x^i d$, for $i$th cylinder in $x$ direction, and $\Delta d_y^i = \gamma_y^i d$ for $i$th cylinder in $y$ direction. Here $\gamma_{x,y}^i \in [-1,1]$ is a random number, and $d$ is the corresponding disorder amplitude. For size disorder, cylinder radii are given by $\Delta r^i = \gamma^i d$, for $i$th cylinder. The largest disorder amplitude, which causes the two nearest-neighbor air-holes to overlap is $d = 0.1a$. We choose the slab of Fig. 2(b) to study the disorder effects, since this slab symmetry gives the best superlensing.

Fig. 4 shows propagation maps for the two kinds of disorder, with the disorder amplitude $d = 0.04a$. Calculated radiation intensity, along the image center parallel to the slab surface, for different disorder amplitudes and types, is shown in Fig. 5. Figs. 4a and 5a are for the position, and the Figs. 4b and 5b are for the size disorder. Disorder affects the propagation maps, but not as severely as one might expect. With an increasing disorder amplitude $d$, the radiation intensity at the image center diminishes slowly with $d$, for $d \leq 0.03a$ (for position disorder) and $d \leq 0.05a$ (for size disorder), followed by a rapid reduction for larger $d$. In addition, there are increasing fluctuations in the position of the image maxima, as well as clear increase in the asymmetry of the image profile, as $d$ increases. These image distortions occur for both kinds of disorder, in roughly the same degree. We define a *good* image as that with the amplitudes of the side peaks along the image plane parallel to the slab surface, being less than10% of the main peak amplitude. With this definition, we conclude that the image quality remains good for the disorder amplitudes $d < 0.05a$, for both kinds of disorder. For larger disorder amplitudes, the image quality rapidly deteriorates. To confirm this statistically, we have calculated 30 randomly generated configurations for the position disorder with $d = 0.07a$, and found that 80% of these configurations have bad image. Therefore, we conclude that disorder does not affect the image quality as long as the amplitude is below 5%. This is roughly in agreement with an earlier work [30], were restricted lensing in a square 2DPC was studied in the presence of only position disorder, and were largely unaffected image remained for disorder amplitudes < 8%.

In conclusion, we study here the superlensing in a slab of a hexagonal 2D-PC, in the presence of differently defined slab surfaces. We find that highly symmetric surfaces minimize the back reflections, leading to improved quality of the imaging. We also study effects of the position and size disorder on the superlensing. We show, that a uniformly disordered photonic crystal superlenses without a significant image deterioration for a disorder amplitude below 5%, for both types of disorder. Since the maximum disorder in our system is 10%, this implies that manufacturing defects will have a minor effect on the performance of the superlenses made by using our system.

## Acknowledgements

This work was supported by US Army Research Development and Engineering Command, Natick Soldier Center, under the grand DAAD16-03-C-0052.

# References

1. V. G. Veselago, Sov. Phys. Ups. **10**, 509 (1968).
2. H. Kosaka, T. Kawashima, A. Tomita, M. Notomi, T. Tamamura, T. Sato, and S. Kawakami, Phys. Rev. B **58**, 10 096 (1998).
3. J.B. Pendry, A.J. Holden, W.J. Stewart, and I. Youngs, Phys. Rev. Lett. 76, 4773 (1996).
4. J.B. Pendry, A.J. Holden, D.J. Robbins, and W.J. Stewart, IEEE Trans. Microwave Theory Tech. 47, 2075 (1999).
5. D.R. Smith, W.J. Padilla, D.C. Vier, S.C. Nemat-Nasser, and S. Schultz, Phys. Rev. Lett. **84**, 4184 (2000).
6. D.R. Smith and N. Kroll, Phys. Rev. Lett. 85, 2933 (2000).
7. R.A. Shelby, D.R. Smith, S.C. Nemat-Nasser, and S. Schultz, Appl. Phys. Lett. **78**, 489 (2001).
8. R.A. Shelby, D.R. Smith, and S. Schultz, Science **292**, 77 (2001).
9. R.W. Ziolkowski and E. Heyman, Phys. Rev. E **64**, 056625 (2001).
10. M. Bayindir, K. Aydin, E. Ozbay, P. Markos, and C.M. Soukoulis, Appl. Phys. Lett. **81**, 120 (2002).
11. R.M. Walser, A.P. Valanju, and P.M. Valanju, Phys. Rev. Lett. **87**, 119701 (2001).
12. D.R. Smith, D. Schurig, and J.B. Pendry, Appl. Phys. Lett. **81**, 2713 (2002).
13. J. B. Pendry, Phys. Rev. Lett. **85**, 3966 (2000).
14. R. Merlin, Appl. Phys. Lett. 84, 1290 (2004).
15. David R. Smith, David Schurig, Marshall Rosenbluth, and Sheldon Schultz, S. Anantha Ramakrishna and John B. Pendry, Appl. Phys. Lett. 82, 1506 (2003).
16. G. Shvets, Phys. Rev. B 67, 035109 (2003).
17. L. V. Panina, A. N. Grigorenko, and D. P. Makhnovskiy, "Optomagnetic composite medium with conducting nanoelements," Phys. Rev. B 66 155411 (2002).

18. V. A. Podolskiy, "Plasmon modes and negative refraction in metal nanowire composites," Optics Express 11, 735-745 (2003), http://www.opticsexpress.org/abstract.cfm?URI=OPEX-11-7-735.
19. M. Notomi, Phys. Rev. B 62, 10 696 (2000).
20. C. Luo, S.G. Johnson, J.D. Joannopoulos, and J.B. Pendry, Phys. Rev. B 65, 201104(R) (2002).
21. C. Luo, S.G. Johnson, and J.D. Joannopoulos, Appl. Phys. Lett. 81, 2352 (2002).
22. Zhi-Yuan Li and Lan-Lan Lin, Phys. Rev. B **68**, 245110 (2003).
23. X. Wang, Z. F. Ren, and K. Kempa, "Unrestricted superlensing in a triangular two dimensional photonic crystal," Opt. Express 12, 2919-2924 (2004), http://www.opticsexpress.org/abstract.cfm?URI=OPEX-12-13-2919
24. J. D. Joannopoulos, R. D. Meade, and J. N. Winn, *Photonic Crystals: Molding the Flow of Light* (Princeton University Press, Princeton, 1995).
25. K. S. Yee, IEEE Trans. Antennas Propag. **14**, 302 (1966).
26. A.Taflove, Computational Electrodynamics—The Finite-Difference Time-Domain Method (Artech House, Norwood, MA, 1995).
27. J. Berenger, J. Comput. Phys. **114**, 185 (1994).
28. S. Foteinopoulou and C. M. Soukoulis, Phys. Rev. B **67**, 235107 (2003).
29. P.V. Parimi, W.T. Lu, P. Vodo, J. Sokoloff, J.S. Derov and S. Sridhar, Phys. Rev. Lett. 92, 127401 (2004).
30. Bikash C. Gupta and Zhen Ye, J. Appl. Phys. 94, 2173 (2003).

# Figure Captions

**Fig. 1**. (a) Calculated photonic crystal band structure for a hexagonal 2DPC of air-holes. Insert shows a circular equi-frequency contour. (b) Effective refractive index of the PC versus frequency.

**Fig. 2.** Calculated electric field patterns generated by oscillating line sources of the electromagnetic waves (along z-direction), for three different slabs of the 2DPC of Fig. 1. The line sources are located in the middle of the dark blue spots on the left sides of the slabs. The vertical scale bar, to the right, quantifies the relative intensity.

**Fig. 3**. Calculated radiation intensity across the image of a line source, parallel to the slab edge, for three slabs of the 2DPC of Fig. 1. Each curve corresponds to a slab shown in Fig. 2.

**Fig. 4**. Calculated electric field patterns generated by oscillating line sources of the electromagnetic waves for two different slabs of the 2DPC of Fig. 1. (a) is for the position, and (b) for the size disorder. Line sources are located in the middle of the dark blue spots on the left sides of the slabs. The horizontal scale bar, in the middle, quantifies the relative intensity.

**Fig. 5**. Calculated radiation intensity across the image, parallel to the slab edge for different disorder amplitudes and types. (a) is for the position, and (b) for the size disorder.

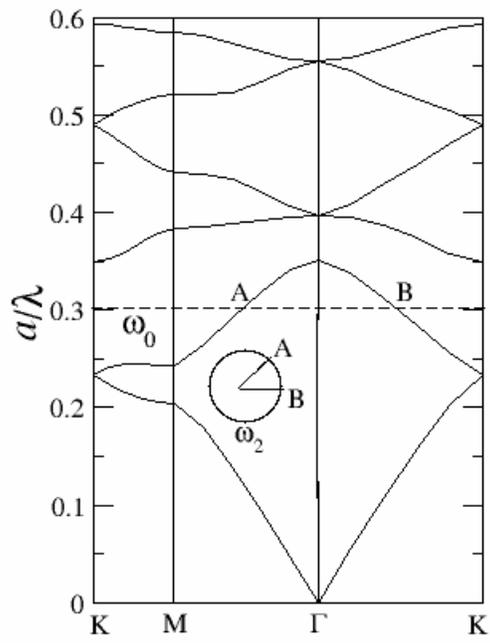

**Fig. 1a**

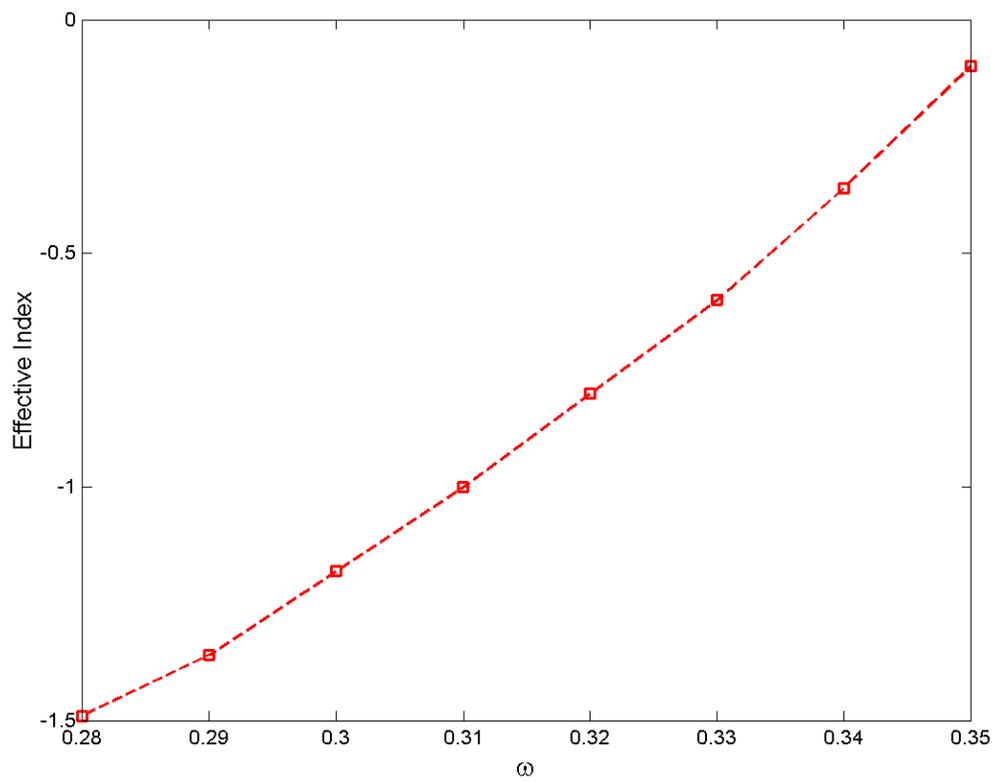

**Fig. 1b**

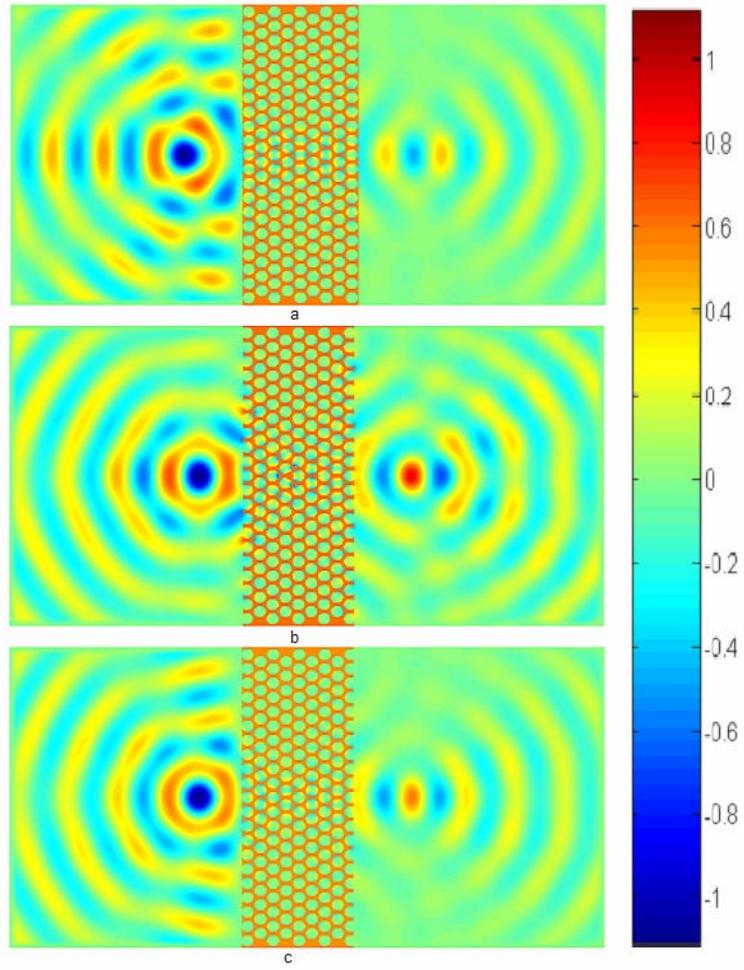

**Fig. 2**

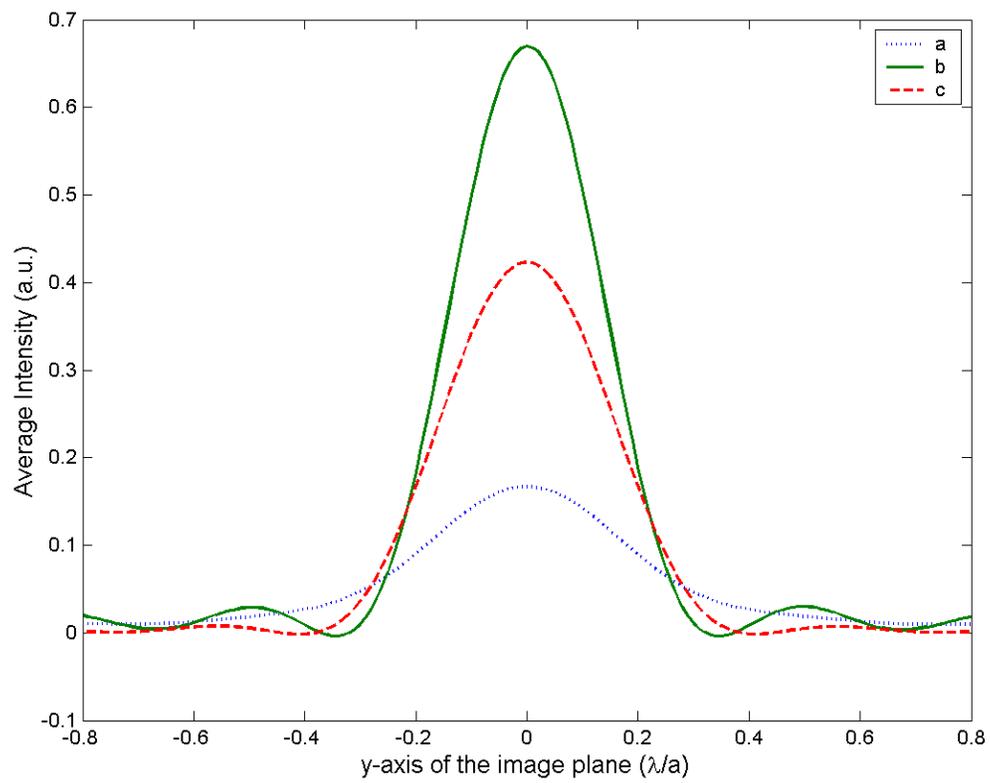

**Fig. 3**

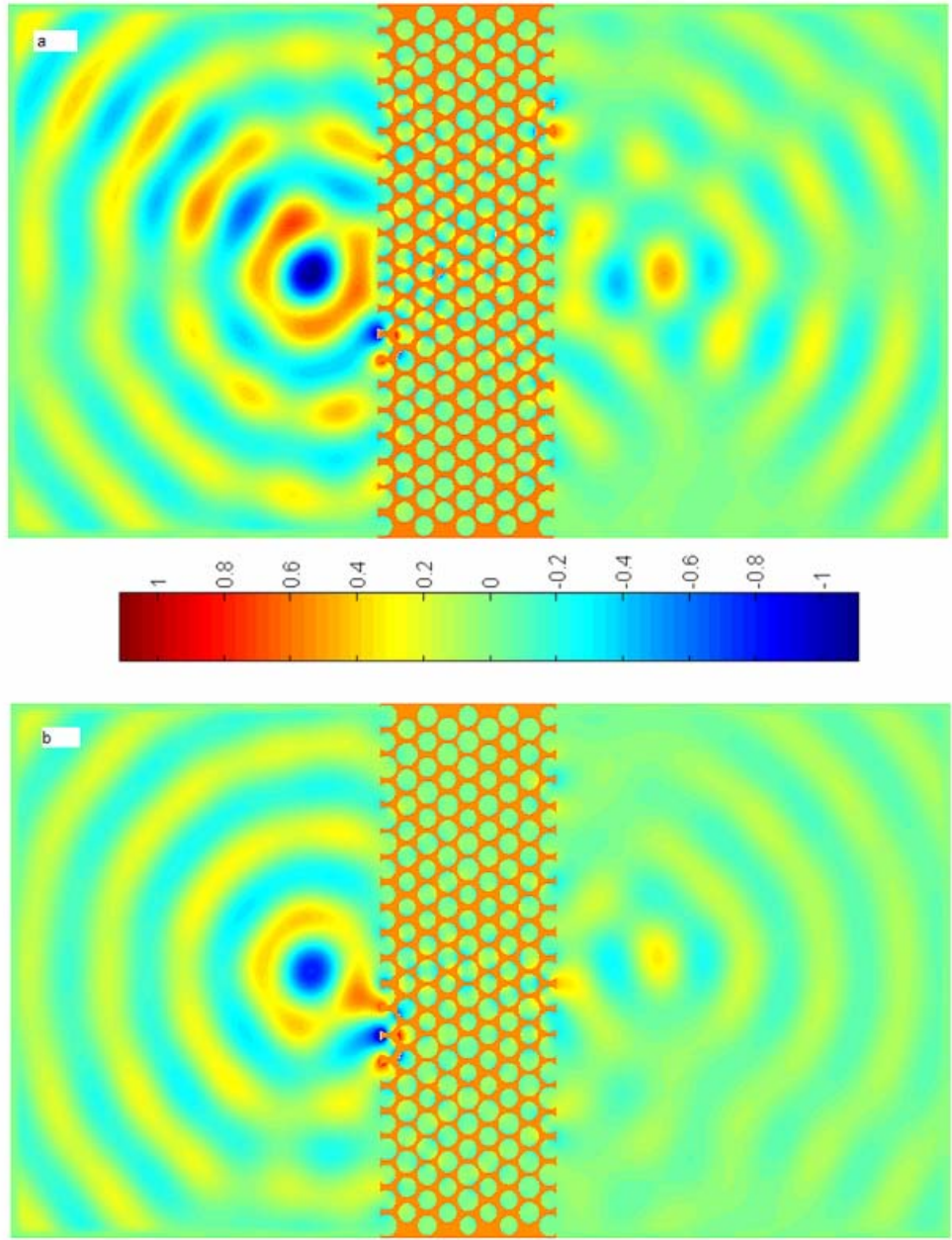

**Fig. 4**

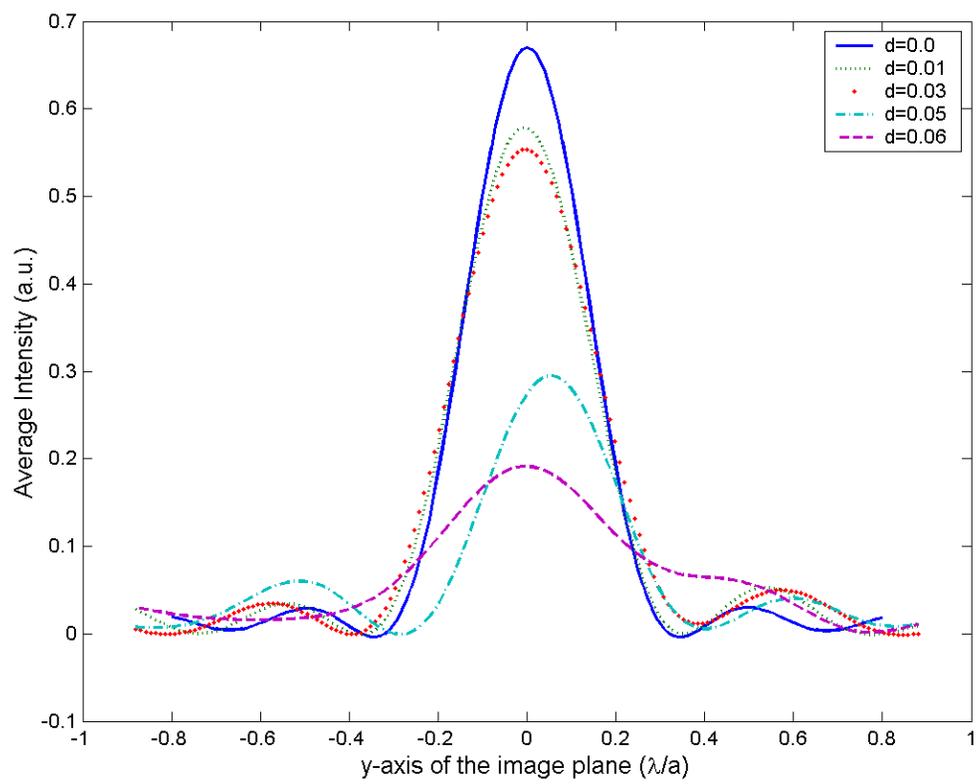

**Fig. 5a**

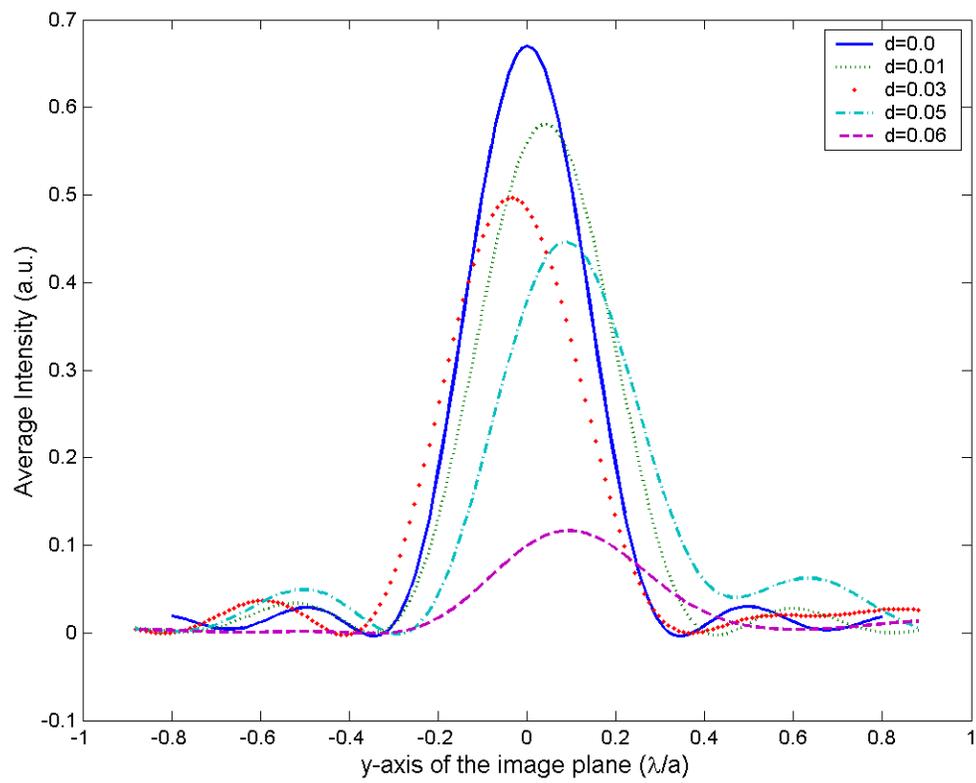

**Fig. 5b**